\documentclass[aps,pre,11pt,a4paper,byrevtex,showpacs,bibnotes,showkeys,showpacs,longbibliography,notitlepage]{revtex4-1}
\usepackage{microtype}
\usepackage{graphicx}
\usepackage{amsmath}
\usepackage{amssymb}
\usepackage{color}
\definecolor{myblue}{rgb}{0.153,0.322,0.706}
\usepackage[colorlinks,linkcolor=myblue,urlcolor=myblue,citecolor=myblue]{hyperref}

\newcommand{\be}{\begin{equation}}
\newcommand{\ee}{\end{equation}}
\newcommand{\ra}{\rightarrow}
\newcommand{\reals}{\mathbb{R}}
\newcommand{\naturals}{\mathbb{N}}
\newcommand{\cL}{\mathcal{L}}
\newcommand{\cH}{\mathcal{H}}
\newcommand{\cW}{\mathcal{W}}

\newcommand{\idf}{1\!\! 1}
\newcommand{\hX}{\hat X}
\newcommand{\sqa}{{\sqrt \alpha}}
\newcommand{\ea}{\frac{\nu}{2}}
\newcommand{\eap}{\ea + \frac{1}{4}}
\newcommand{\Diff}[1]{\partial_{x}{#1}}
\newcommand{\hF}[5]{{ _{#1\!}F_{#2}\p{#3;#4;#5}}}
\newcommand{\hM }[3]{ \hF 1 1 {#1} {#2} {#3}} 
\newcommand{\p}[1]{\left(#1\right)}
\newcommand{\id}{\mathbb I}

\begin{document}
\title{Diffusions conditioned on occupation measures}

\author{Florian Angeletti}
\affiliation{National Institute for Theoretical Physics (NITheP), Stellenbosch 7600, South Africa}
\affiliation{Institute for Theoretical Physics, Department of Physics, Stellenbosch University, Stellenbosch 7600, South Africa}

\author{Hugo Touchette}
\email{htouchette@sun.ac.za}
\affiliation{National Institute for Theoretical Physics (NITheP), Stellenbosch 7600, South Africa}
\affiliation{Institute for Theoretical Physics, Department of Physics, Stellenbosch University, Stellenbosch 7600, South Africa}
\date{\today}

\begin{abstract}
A Markov process fluctuating away from its typical behavior can be represented in the long-time limit by another Markov process, called the effective or driven process, having the same stationary states as the original process conditioned on the fluctuation observed. We construct here this driven process for diffusions spending an atypical fraction of their evolution in some region of state space, corresponding mathematically to stochastic differential equations conditioned on occupation measures. As an illustration, we consider the Langevin equation conditioned on staying for a fraction of time in different intervals of the real line, including the positive half-line which leads to a generalization of the Brownian meander problem. Other applications related to quasi-stationary distributions, metastable states, noisy chemical reactions, queues, and random walks are discussed.
\end{abstract}



\maketitle

\section{Introduction}

The stationary distribution of a Markov process gives, when it is unique, the average fraction of time the process spends in any given state in the long-time limit. When finite-time trajectories are considered, fluctuations around this average occupation occur, with a probability that depends on the forces and noise acting on the process. The position of a Brownian particle, for example, is positive in one dimension on average half of the time, yet sample trajectories have a strong tendency to stay positive or negative for any finite time, pushing the positive occupation above or below~$\frac{1}{2}$. Similarly, Brownian particles evolving in complex potentials tend to spend most of their time around the stable equilibria of the acting potential, but are also likely to `climb' it in finite time to reach possible unstable or metastable states.

Similar fluctuations of the occupation that persist in time are observed in almost all random systems, including jump processes describing noisy chemical reactions and particle transport \cite{kampen1992,gardiner1985,derrida2007,sekimoto2010}, phase ordering and coarsening dynamics in magnetic systems \cite{dornic1998,drouffe1998,baldassarri1999}, financial time series \cite{bouchaud2000b}, queueing systems \cite{shwartz1995}, as well as random walks on graphs \cite{montanari2002,kishore2012,kishore2013,bacco2015}. In these and many other applications, it is of interest not only to determine the probability that a process ventures in an atypical region of the state space, for example, around a metastable or unstable state, but also to describe with a \emph{modified process} the effective dynamics of the process in that region. 

\begin{figure*}[t]
\includegraphics[width=2.9in]{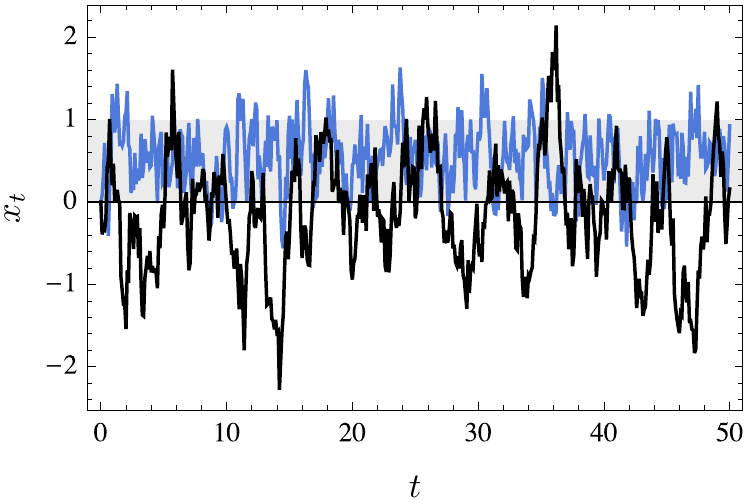}%
\hspace{0.5in}%
\includegraphics[width=2.9in]{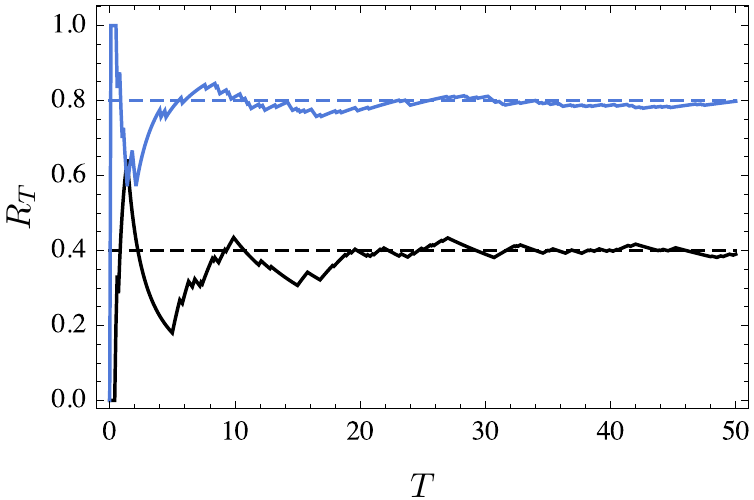}
\caption{Illustration of the driven process for the occupation region $S=[0,1]$. Left: Sample trajectory of a process $X_t$ (black curve) spending about $40\%$ of its time in $S$ (gray region) compared to a sample trajectory of the driven process $\hX_t$ (blue curve) representing the process $X_t$ conditioned on spending $80\%$ of its time in $S$. Right: Fraction $R_T$ of the time interval $[0,T]$ spent in $S$ as a function of $T$ for $X_t$ (black) and $\hX_t$ (blue).}
\label{figtraj1}
\end{figure*}

We show in this paper how to formulate this problem as a Markov occupation conditioning problem which can be solved using the general framework proposed recently in \cite{chetrite2013,chetrite2014,chetrite2015}. The general idea is illustrated in Fig.~\ref{figtraj1}. We consider a general Markov process $X_t$ and condition probabilistically that process on spending a fraction $R_T$ of the time interval $[0,T]$ in some subset $S$ of its state space. Following \cite{chetrite2013,chetrite2014,chetrite2015}, we then derive a new Markov process $\hX_t$, called the \emph{driven} process, which is equivalent to the conditioned process at the level of stationary states. In particular, the mean occupation of $\hX_t$ in $S$ is $R_T$, so it realizes what is a fluctuation for $X_t$ in a typical way. 

The driven process is in this sense the modified process mentioned before: it represents the effective (stochastic) dynamics of $X_t$ as this process is seen to `fluctuate' in $S$ for a fraction of time given by the occupation measure $R_T$. When applied to noisy chemical reactions, for example, the driven process gives an effective chemical reaction with modified rates accounting for concentration fluctuations. For a random walk visiting a `rare' graph component, it gives a new random walk that concentrates on that component.

This effective representation of fluctuations can be constructed for any ergodic Markov processes, including Markov chains and jump processes. Here, we focus on diffusions described by stochastic differential equations in order to provide a new application of the framework developed in \cite{chetrite2013,chetrite2014,chetrite2015} and to set a template for applications based on continuous-space and continuous-time Markov models. As an example, we study in Sec.~\ref{secOU} the Langevin equation conditioned on staying in the interval $[a,b]$, the half-line $[a,\infty)$, and the point $\{a\}$ for a fraction of the time interval $[0,T]$. The second conditioning is a variant of the so-called \emph{Brownian meander}, corresponding to Brownian motion restricted to stay positive for $t\in [0,T]$ \cite{revuz1999,janson2007,majumdar2015}. Other physical and manmade applications of the driven process related to more complex diffusions, jump processes, and random walks are mentioned in the conclusion of the paper.

\section{Occupation conditioning}
\label{secOC}

We explain in this section how the conditioned and the driven processes are constructed for a conditioning involving an occupation measure. This is a special case of the framework presented in \cite{chetrite2013,chetrite2014,chetrite2015} dealing with general, time-integrated random variables for the conditioning. 

\subsection{Model}

We consider a pure diffusion process $X_t\in\reals^m$ described by the following (It\^o) stochastic differential equation (SDE):
\be
dX_t=F(X_t)dt+\sigma dW_t,
\label{eqsde1}
\ee
where $F:\reals^m\ra\reals^m$ is the drift, $W_t$ is an $n$-dimensional Brownian motion, and $\sigma$ is the $m\times n$ noise matrix, assumed for simplicity to be constant in space and non-singular (invertible) \footnote{See \cite{chetrite2014} for the case of multiplicative noise involving a noise matrix $\sigma(x)$ depending on $X_t$.}. The probability density $p(x,t)$ of this process evolves according to the Fokker-Planck equation
\be
\partial_t p(x,t)=L^\dag p(x,t),
\ee
expressed here in terms of the Fokker-Planck operator,
\be
L^\dag =-\nabla \cdot F +\frac{1}{2}\nabla\cdot D\nabla
\ee
with $D=\sigma\sigma^T$ the diffusion matrix. For the remaining, we also need the adjoint of the Fokker-Planck operator, 
\be
L=F\cdot \nabla+\frac{1}{2}\nabla\cdot D\nabla,
\label{eqgen1}
\ee
which generates the evolution of expectations of $X_t$ \cite{risken1996}.

Given the evolution of $X_t$, we now consider a subset $S\subset\reals^m$ and look at the fraction of time that $X_t$ spends in $S$ in the time interval $[0,T]$ using 
\be
R_{T}=\frac{1}{T}\int_0^T \idf_S(X_t)\, dt
\label{eqempmeas1}
\ee
where $\idf_S(x)$ denotes the indicator function equal to $1$ if $x\in S$ and $0$ otherwise. This random variable, which explicitly depends on both $S$ and $T$, is called the \emph{occupation measure} of $S$ or the (normalized) \emph{local time} when $S$ is a single point.  Assuming that $X_t$ has a unique stationary distribution $p^*$ satisfying $L^\dag p^*=0$, we have by the ergodic theorem that 
\be
\lim_{T\ra\infty} R_T =E_{p^*}[\idf_S]=\int_S p^*(x)\, dx,
\ee
so that $R_T$ converges in probability to the mean occupation in $S$ given by $p^*(S)$. 

For finite integration times, $R_T$ fluctuates around this concentration point according to its probability density $P(R_T=r)$, which can be expressed for large times as 
\be
P(R_{T}=r) = e^{-T I(r)+o(T)}
\label{eqldp1}
\ee
or, equivalently,
\be
\lim_{T\ra\infty} -\frac{1}{T}\ln P(R_T=r)=I(r).
\ee
This scaling of the distribution is known as a \emph{large deviation principle} (LDP) \cite{dembo1998,hollander2000,touchette2009}. The rate of decay $I(r)$ is called the \emph{rate function} and can be obtained from the \emph{contraction principle} of large deviation theory by the following minimization:
\be
I(r)=\min_{\rho:C(\rho)=r} J(\rho),
\label{eqcps1}
\ee
which involves the Donsker-Varadhan or level-2 rate function,
\be
J(\rho)=-\min_{h>0} \int \rho(x) (h^{-1} Lh)(x)\, dx,
\label{eqdv1}
\ee
and the contraction linking $\rho$ to $R_T$:
\be
C(\rho)=\int_{\reals^m} \rho(x) \idf_S(x)\, dx=\int_S \rho(x) \, dx.
\ee
This result is derived in Appendix~\ref{appLDP1}.

The LDP (\ref{eqldp1}) shows that $X_t$ is exponentially unlikely for long times $T$ to enter the region $S$ for a fraction $R_T$ of time, except when $R_T$ is the stationary fraction $r^*$ of time spent in $S$. The ergodic theorem indeed states that  $P(R_T=r^*)\ra 1$ as $T\ra\infty$, which implies $I(r^*)=0$, corresponding to the \emph{typical} occupation of $X_t$. Any other fraction $R_T\neq r^*$ represents an \emph{atypical} occupation of $X_t$ in $S$ characterized by $I(r)>0$ and so $P(R_T=r)\ra 0$ as $T\ra\infty$. For more information about the large deviations and applications of occupation times, see \cite{majumdar2002b,majumdar2002c,majumdar2005,sabhapandit2006}

\subsection{Conditioned and driven processes}

We now consider a fixed occupation $R_T=r$ of $X_t$ in $S$ and derive the effective driven process $\hX_t$ that describes $X_t$ conditioned on (or restricted to) this occupation. The construction of $\hX_t$ is explained in \cite{chetrite2013,chetrite2014,chetrite2015} and requires that we find the dominant eigenvalue $\lambda(k)$ and corresponding eigenfunction $r_k$ of the \emph{tilted generator}, defined by
\be
\cL_k=L+k\idf_S,
\ee
where $k\in\reals$ and $L$ is the generator (\ref{eqgen1}) of $X_t$. With these elements, the driven process is defined as the Markov process with modified generator
\be
L_k= r_k^{-1} \cL_k r_k -r_k^{-1}(\cL_kr_k) 
\label{eqdoob1}
\ee
acting on functions $f$ according to
\begin{eqnarray}
(L_kf)(x) &=&\frac{1}{r_k(x)} (\cL_k r_k f)(x)-\frac{1}{r_k(x)} (\cL_k r_k)(x) f(x)\nonumber\\
&=& \frac{1}{r_k(x)} (\cL_k r_k f)(x)-\lambda(k) f(x).
\end{eqnarray}
As shown in \cite{chetrite2014}, the effect of this transform on the SDE (\ref{eqsde1}) is to change the drift $F$ to the modified or \emph{driven} drift
\be \label{eq:effective_f}
F_k(x)=F(x)+D\nabla\ln r_k(x)
\ee
while keeping the diffusion matrix $D$ constant. The evolution of the driven process $\hX_t$ is thus given by the modified SDE 
\be
d\hX_t=F_k(\hX_t)dt+\sigma dW_t
\ee
perturbed by the same Gaussian noise as $X_t$ but involving the new driven drift $F_k$.

The connection between the driven process and the conditioning of $X_t$ on $R_T=r$ is illustrated again in Fig.~\ref{figtraj1} and is fully explained in \cite{chetrite2014}. The idea briefly is that the driven process $\hX_t$ and the conditioned process $X_t|R_T=r$ have the same stationary properties, in addition to having similar probabilities for their trajectories as $T\ra\infty$,  if the rate function $I(r)$ of $R_T$ is convex and $k$ is chosen so that 
\be
k=I'(r).
\label{eqtemp1}
\ee
In this sense, we then say that $\hX_t$ is \emph{equivalent} to $X_t|R_T=r$ or \emph{realizes} that conditioned process in the long-time limit. 

This equivalence is similar to the equivalence of the microcanonical and canonical ensembles in equilibrium statistical mechanics \cite{chetrite2014}: the conditioned process $X_t|R_T=r$ is essentially a process generalization of the microcanonical ensemble in which the `energy' $R_T$ is constant and equal to $r$, whereas the driven process is a generalization of the canonical ensemble in which $R_T$ fluctuates but concentrates in the `thermodynamic limit' $T\ra\infty$ to $r$, the constant of the microcanonical ensemble. This is achieved by matching the `temperature' $k$ to the constraint $R_T=r$ according to (\ref{eqtemp1}), which is an analog of the thermodynamic temperature-energy relation.

Another way to understand the driven process is as an optimal change of measure or process \cite{chetrite2015}. Recall that the event $R_T=r$ is a rare fluctuation in the original process $X_t$ having an exponentially small probability for long times $T$. The driven process, by contrast, is such that $R_T=r$ happens with certainty as $T\ra\infty$, so that the transformation (\ref{eqdoob1}) modifies the process $X_t$ to make a rare occupation typical. In general, many transformed processes can be used to achieve this reweighting of rare events. The driven process is special in that it the process \emph{closest} to $X_t$, with respect to a metric defined by the relative entropy, that makes the occupation $R_T=r$ typical; see \cite{chetrite2015} for more details.

\subsection{Spectral problem and effective potential}

The difficulty of constructing the driven process comes from solving the spectral problem
\be 
\cL_kr_k=\lambda(k) r_k
\label{eq:spectral:right}
\ee
for the dominant eigenvalue and its corresponding eigenfunction. Depending on the form of generator $L$ considered and, more precisely, its self-adjointness, three cases arise:
\begin{description}
\item[Case 1] $L=L^\dag$. This is the simplest case determining a reversible process  with respect to the Lebesgue (uniform) measure. In this case, the techniques of quantum mechanics apply: the spectrum of $L$ or $\cL_k$ is real and the eigenfunction $r_k$ must be found by solving (\ref{eq:spectral:right}) with vanishing boundary condition for $r_k^2(x)$ as $|x|\ra\infty$.

\item[Case 2] $L\neq L^\dag$ but the spectrum of $L$ is real. This arises, for example, when $X_t$ is a reversible or equilibrium diffusion having a gradient drift 
\be
F=-\frac{D}{2}\nabla U
\label{eq:force-potential}
\ee 
and, therefore, a Gibbs stationary distribution
\be
p^*(x)=e^{-U(x)}.
\ee
In this case, it is known that $L$ is self-adjoint with respect to an inner product defined with $p^*$ and that this can be used to `symmetrize' $L$ into a self-adjoint operator $H$, playing the role of a quantum Hamiltonian \cite{majumdar2002}. This symmetrization is simply defined as 
\be
H=e^{-U/2}Le^{U/2}
\ee 
and leads, when applied to $\cL_k$, to the \emph{tilted Hamiltonian}
\be
\cH_k =e^{-U/2} \cL_k e^{U/2}
=\frac{D}{2} \left[ \Delta  + \frac{\Delta  U}{2} -  \left(\frac{\nabla  U }{2}\right)^2 \right] + k \idf_S. 
\label{eq:hamiltonian}
\ee

This operator has of course the same real spectrum as $\cL_k$, so that
\be
\cH_k\psi_k=\lambda(k)\psi_k,
\ee
but its dominant eigenfunction $\psi_k$, obtained with the natural vanishing boundary condition $\psi_k(x)^2=0$ at infinity, is related to $r_k$ by $r_k=e^{U/2}\psi_k$.

\item[Case 3] $L\neq L^\dag$ and the spectrum is complex. This happens when $F$ is not gradient, $\sigma$ depends on $X_t$, or external reservoirs are included in this process as boundary conditions. In this case, $X_t$ represents a genuine nonequilibrium process supporting non-vanishing probability currents, for which $L$ or $\cL_k$ cannot be symmetrized. Moreover, the spectral problem (\ref{eq:spectral:right}) on its own is incomplete: it must be solved in tandem with the dual problem
\be 
\cL_k^\dag l_k=\lambda(k) l_k
\label{eq:spectral:left}
\ee
and the the boundary condition $l_k(x)r_k(x)=0$ at infinity \footnote{This arises from the definition of the adjoint with the inner product based on the Lebesgue measure.}. This is more difficult to solve in general than the case of self-adjoint operators (Cases~1 and~2).
\end{description}

We focus in the rest of the paper mostly on Case 2, which is equivalent to a quantum ground state problem with effective Schr\"odinger  Hamiltonian $\cH_k$. Assuming that the drift is conservative, as in (\ref{eq:force-potential}), we can express the driven drift (\ref{eq:effective_f}) in this case also in gradient form,
\be
F_k=-\frac{D}{2}\nabla U_k,
\label{eq:eff-potential}
\ee
by introducing the \emph{effective} or \emph{driven potential} 
\be
U_k(x)=U(x)-2\ln r_k(x) =-2\ln \psi_k(x),
\label{eq:eff-potential-eigen} 
\ee
which realizes the occupation conditioning.

Non-reversible diffusions falling in Case 3 cannot be represented by such an effective potential, even though the modified drift $F_k$, as given by (\ref{eq:effective_f}), is always a gradient perturbation of the original drift $F$ when $D$ is constant. This property of $F_k$ comes from the time-additive form of $R_T$. For other conditionings, based for example on currents or the entropy production, the perturbation $F_k-F$ can have a non-conservative and, therefore, nonequilibrium component; see Sec.~5.5 of \cite{chetrite2014} for more detail.

\section{Application}
\label{secOU}

We illustrate in this section our results for an exactly-solvable model based on the linear Langevin equation or one-dimensional Ornstein-Uhlenbeck process defined by
\be
dX_t=-\gamma X_tdt+\sigma dW_t,
\label{eq:ou1}
\ee
where $X_t\in\reals$, $W_t\in\reals$, with $\gamma$ and $\sigma$ positive constants. This process obviously falls in Case 2 of the previous section, as do all processes defined on $\reals$ without sinks or sources. The linear drift $F(x)=-\gamma x$ is associated with the parabolic potential
\be
U(x)=\frac{\alpha x^2}{2},
\ee
where $\alpha =2\gamma/ D$ and $D=\sigma^2$. The quantum problem that we need to solve therefore is 
\be 
\left[ \frac{d^2}{dx^2}   + \frac{\alpha }{2} - \frac{\alpha ^2}{4} x^2 + \frac{2k}{D} \idf_S(x)  \right] \psi(x)  = \frac{2 \lambda}{D} \psi(x),
\label{eq:eigen:chi}
\ee
which we convert to 
\be  
\Psi''(x)   - \frac{x^2}{4} \Psi(x) +\left[ \frac{1}{2} + \frac{2}{D \alpha}\Big( k \idf_{\sqa S}(x) - \lambda \Big)\right] \Psi(x)  =0
\label{eq:eigen:chi:adim}
\ee
with the rescaling $\Psi (x) = \psi(x/\sqa)$. The same quantum problem can be obtained using path integral methods as applied to Brownian functionals and the Feynman-Kac equation; see \cite{majumdar2002,majumdar2005}.

Equation~(\ref{eq:eigen:chi:adim}) is essentially the Weber equation with piecewise-constant coefficients, representing a quantum harmonic oscillator with piecewise-shifted potential. It can be solved exactly in and out of the conditioning interval $S$ and then pieced together by requiring continuity at the boundaries of $S$. This is done next for three occupations, namely, $S=[a,b]$ (finite interval), $S=[a,\infty)$ (half-line), and $S=\{a\}$ (point conditioning).

\subsection{Finite interval}

For $S=[a,b]$, we must solve (\ref{eq:eigen:chi:adim}) on the three regions $(-\infty , a)$, $[a,b]$, and $(b, \infty)$, and piece the three solutions, as mentioned, continuously at $x=a$ and $x=b$. Over each region, the Weber equation has the form
\be 
\Psi''(x)-  \left( \frac{x^2}{4}  + \nu (x) \right)\Psi(x)=0,
\label{eq:eigen:chi:ext}
\ee
where
\be
\nu (x)= \frac{2}{\alpha  D } \left(\lambda  - \frac{\alpha  D} {4} - k \idf _{\sqa S}(x) \right). 
\label{eq:nu1}
\ee
This function takes only two values, denoted from now on by $\nu =\nu (S)$ and $\nu'=\nu(\reals\backslash S)$.

The solution space of the Weber equation is spanned by
\be 
\begin{aligned}
s_1(\nu,x) &= e^{-\frac{x^2}{4} } \, \hM{\frac{\nu }{2} + \frac 1 4}{\frac 1 2} {\frac{x^2}{2} } \\
s_2(\nu,x) &=  x e^{-\frac{x^2}{4} } \, \hM{\frac{\nu }{2} + \frac 3 4}{\frac 3 2} {\frac{x^2}{2} },
\end{aligned} 
\ee
where $\hM{a}{b}{x}$ is the confluent hypergeometric function of the first kind. From these two particular solutions, it is possible to construct a solution outside $[a,b]$ that decays to 0 at infinity:
\begin{widetext}
\be
W( \nu , x ) = \frac 1 { 2^{\eap} \sqrt {\pi } } \left[ \cos \p{\textstyle\ea \pi  + \frac{\pi } 4 } \Gamma \p{\textstyle\frac 1 4 - \ea} s_1(\nu,x) 
- \sqrt {2} \sin\p{\textstyle\ea \pi  + \frac{\pi } 4 } \Gamma \p{\textstyle\frac 3 4 - \ea} s_2(\nu,x)  \right].
\ee
Combining these solutions, we then construct the complete eigenfunction as
\be
\Psi (x) =  
\begin{cases} 
K_1 W( \nu' , -x) & x < a \\
K_2 s_1( \nu , x) + K_3 s_2( \nu , x) & a<x<b \\
K_4 W( \nu' , x ) & b < x, \\
\end{cases}
\label{eq:fullsol1}
\ee
where the $K_i$'s are constants to be adjusted by imposing continuity. 

To this end, we define the vector $K=(K_1,K_2,K_3,K_4)^T$ and the matrix
\be
C(\lambda,k)=
\begin{pmatrix}
 - W( \nu ' , - a) & s_1( \nu , a )   & s_2( \nu , a ) & 0 \\
   \Diff{W}( \nu ', - a) & \Diff{s_1}( \nu , a ) & \Diff{s_2}( \nu , a ) & 0 \\
 0  &  s_1( \nu , b )   & s_2( \nu , b ) & - W( \nu ', b ) \\
 0   &  \Diff{s_1}( \nu , b )   & \Diff{s_2}( \nu , b ) & - \Diff{W}( \nu ', b ) \\
\end{pmatrix},
\ee 
\end{widetext}
where $\partial_{x}$ denotes the first derivative with respect to the second coordinate (noted $x$ above). The continuity of $\Psi(x)$ at $a$ and $b$ is equivalent to the following linear equation:
\be
C(\lambda ,k) \, K = 0.
\ee
Non-trivial solutions therefore exist if, and only if,
\be 
\det C(\lambda , k) =0.
\label{eq:eigen:trans}
\ee
This defines a transcendental equation in $\lambda$ involving hypergeometric functions, which can easily be solved numerically to obtain $\lambda(k)$ with an arbitrary precision. To find the associated rate function $I(r)$, we then use the fact that $\lambda(k)$ and $I(r)$ are related by Legendre transform when the former is differentiable \cite{dembo1998,hollander2000,touchette2009}:
\be
I(r)=\sup_{k\in\reals}\{kr -\lambda(k)\}.
\ee
In parametric form, we therefore have
\be 
 I( \lambda '(k) ) = k \lambda '(k) - \lambda (k),\qquad k\in\reals.
 \label{eq:I:parametric}
\ee

\begin{figure*}[t]
\centering
\includegraphics[width=2.9in]{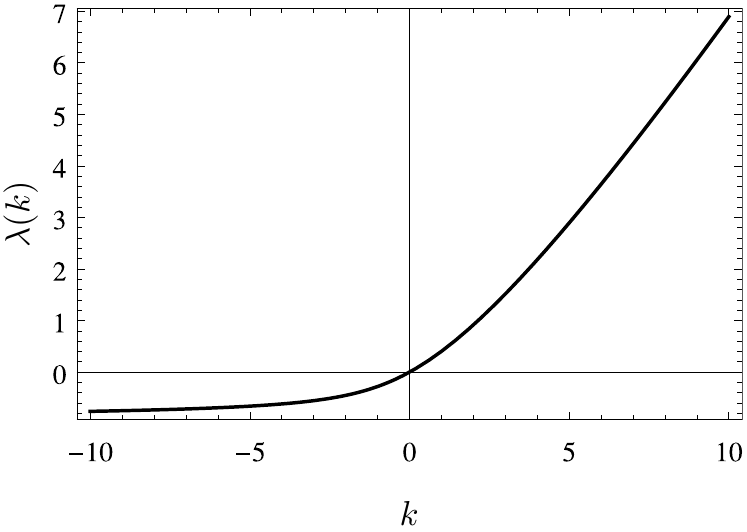}%
\hspace*{0.5in}%
\includegraphics[width=3in]{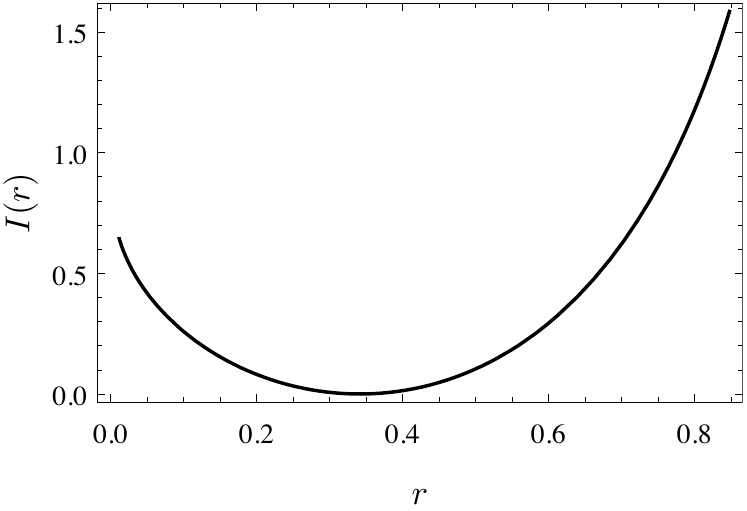} 
\caption{SCGF (left) and rate function (right) for $S=[0,1]$. Parameters: $\alpha=1$, $D=2$.}
\label{fig:rate}
\end{figure*}

Figure~\ref{fig:rate} shows the result of these expressions for $S=[0,1]$. As can be seen, the tails of $\lambda(k)$ are asymptotically linear with slopes $0$ and $1$ as $|k|\ra\infty$, reflecting the fact that $I(r)$ is defined for $r\in [0,1]$ and is steep at $r=0$ and $r=1$. This is important for what follows as it means, following (\ref{eqtemp1}), that the effective potential associated with no occupation in $[a,b]$ is obtained by taking the limit $k\ra -\infty$, whereas full occupation in $[a,b]$ is obtained with $k\ra\infty$. In between, $k$ is related to the occupation fraction $r$ via (\ref{eqtemp1}) or equivalently $\lambda'(k)=r$.

\begin{figure*}[t]
\centering
\includegraphics[width=2.9in]{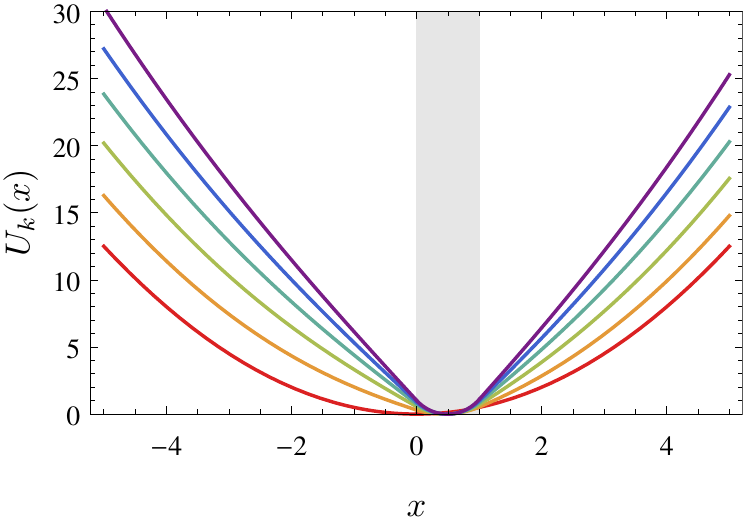}%
\hspace*{0.5in}%
\includegraphics[width=2.9in]{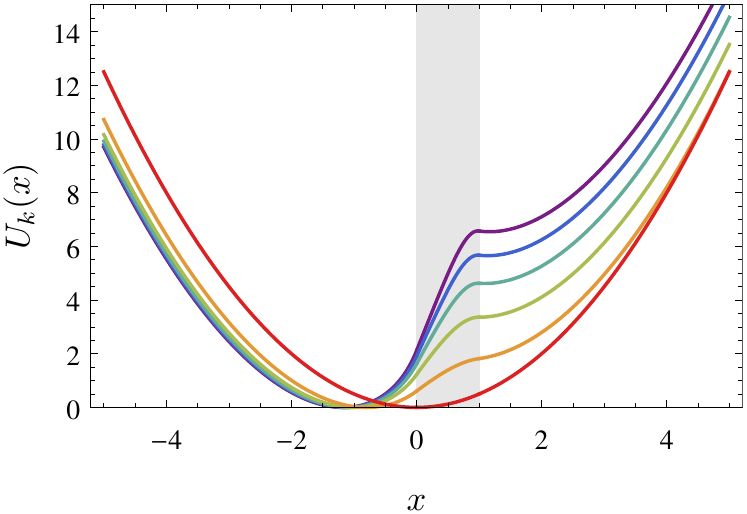}
\caption{(Color online) Effective potential $U_k(x)$ for $S=[0,1]$. Left: $k =0:2:10$ (from bottom to top curves) using the notation $k=k_{\min} : dk : k_{\max}$. Right: $k=0:-2:-10$ (from bottom to top curves). Parameters: $\alpha=1$, $D=2$.}
\label{fig:asym:effective_potential}
\end{figure*}

\begin{figure}[t]
\includegraphics[width=2.9in]{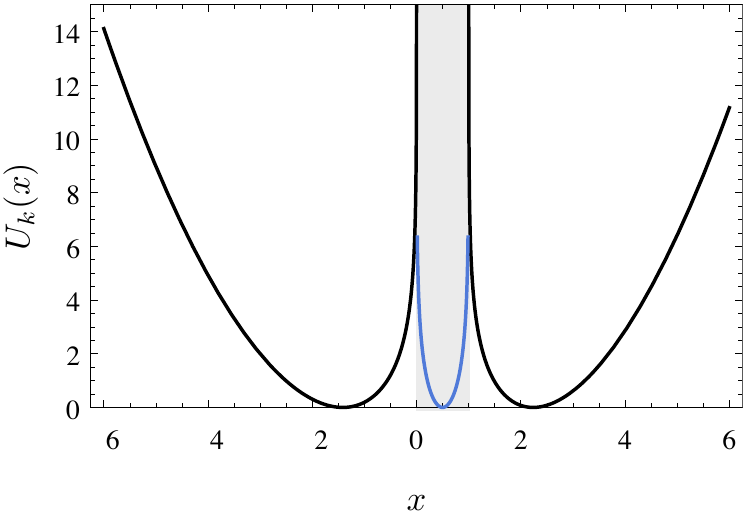}
\caption{(Color online) Black: Effective potential $U_{-\infty}(x)$ preventing any occupation in $S=[0,1]$. Blue: Effective potential $U_\infty(x)$ forcing a total occupation in $S=[0,1]$. Parameters: $\alpha=1$, $D=2$.}
\label{fig:asym:limit_potential}
\end{figure}

To find the effective potential $U_k$, we compute the kernel of the matrix $C(k,\lambda )$ to obtain $\Psi$ via (\ref{eq:fullsol1}), and then rescale $\Psi$ back to $\psi$. Figure~\ref{fig:asym:effective_potential} shows the result of these calculations for $S=[0,1]$ and different values of $k$. For $k>0$, we see that $U_k(x)$ becomes steeper around $[0,1]$ compared to the `natural' potential $U(x)$ obtained for $k=0$. This confines the process inside $[0,1]$, and so increases naturally the time spent inside this interval. In the limit $k\ra\infty$, the process is completely confined inside that interval by an infinitely-steep potential $U_\infty(x)$ shown in Fig.~\ref{fig:asym:limit_potential}. In this case, it is easy to see by analogy with the confined harmonic oscillator \cite{dean1966,consortini1976} that $U_{\infty}(x)$ must diverge logarithmically near $x=0$ and $x=1$, since $\psi_k(x)$ vanishes at these points. This yields a diffusive version of the so-called $Q$-process, arising in the context of quasi-stationary distributions \cite{darroch1965,darroch1967,villemonais2012,collet2014}, which corresponds here to the Ornstein-Uhlenbeck process conditioned on not leaving $[0,1]$.

The effective potential is more interesting for $k<0$. In this case, a non-trivial barrier develops inside $[0,1]$ so as to `deconfine' the process from $[0,1]$, leading to a reduced occupation in that interval. As $k\ra-\infty$, $U_k(x)$ becomes steep near $x=0^-$, as shown in Fig.~\ref{fig:asym:limit_potential}, preventing the process to reach $[0,1]$ from negative initial conditions. It also becomes steep near $x=1^+$ while being raised, as shown in the left plot of Fig.~\ref{fig:asym:effective_potential}. However, because the height of the potential obtained for $x>1$ does not play any role when it becomes disconnected from the one obtained for $x<0$ \footnote{Only the gradient of the potential has a physical meaning.}, we can shift the former down to zero, yielding the limiting potential shown in Fig.~\ref{fig:asym:limit_potential}. This leads effectively to a breaking of ergodicity for the process conditioned on not entering $[0,1]$: the process started in the region $x<0$ stays in that region and cannot visit the region $x>1$ because of the infinite barrier at $x=0$. Conversely, when $X_t$ is started in the region $x>1$, it stays in that region and cannot cross to $x<0$. For initial conditions in $[0,1]$, the process is not defined, at least not in the formal limit $k=-\infty$. For any finite $k$, however, the driven process is ergodic.

\subsection{Half line}

We now consider $S=[a,\infty)$ as the occupation set to show how our results can be used to study variants of the Brownian meander process corresponding to Brownian motion conditioned on staying positive. The Weber solution in this case has two branches:
\be
\Psi (x) =  \begin{cases} 
K_1 W( \nu' ,-x) & x < a \\
K_2 W(\nu , x) &  x > a  \\
 \end{cases}
\ee
linked continuously at $x=a$ by solving (\ref{eq:eigen:trans}) using the matrix
\be
C(\lambda ,k) = 
\begin{pmatrix}
 - W( \nu ', - a) & W( \nu , a)  \\
 \Diff{W}( \nu ', -a) & \Diff{W}( \nu , a)
\end{pmatrix}.
\ee

\begin{figure*}[t]
\centering
\includegraphics[width=2.9in]{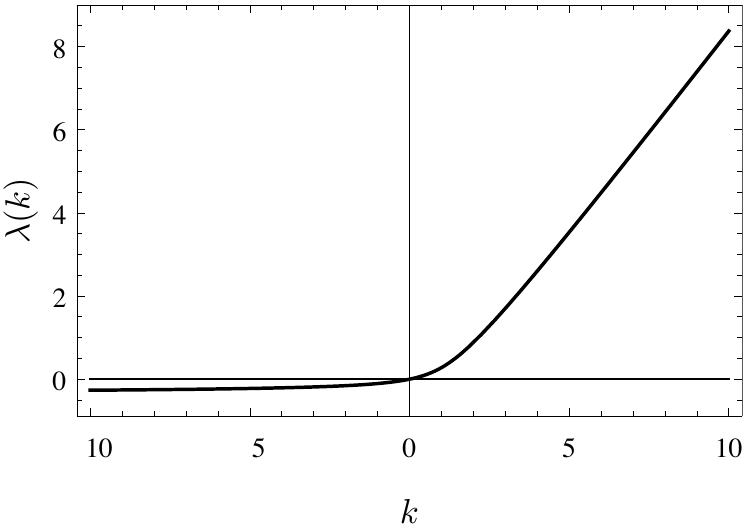}%
\hspace*{0.5in}
\includegraphics[width=3in]{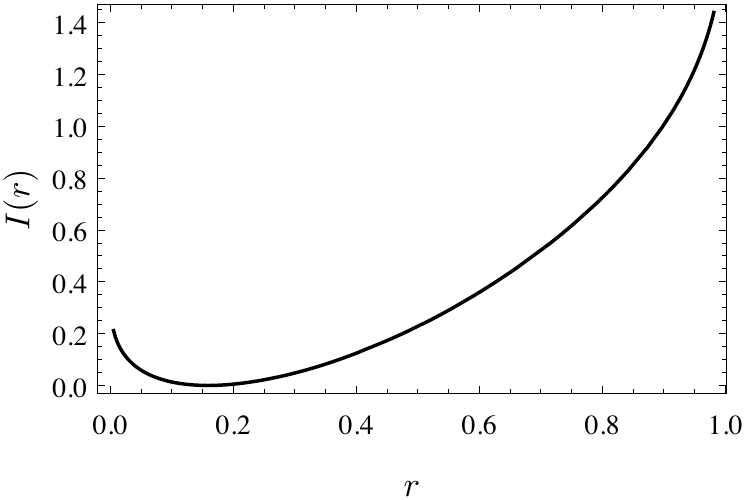}
\caption{SCGF (left) and rate function (right) for $S=[1,\infty)$. Parameters: $\alpha=1$, $D=2$.}
\label{fig:half:lambda}
\end{figure*}

\begin{figure*}[t]
\centering
\includegraphics[width=2.9in]{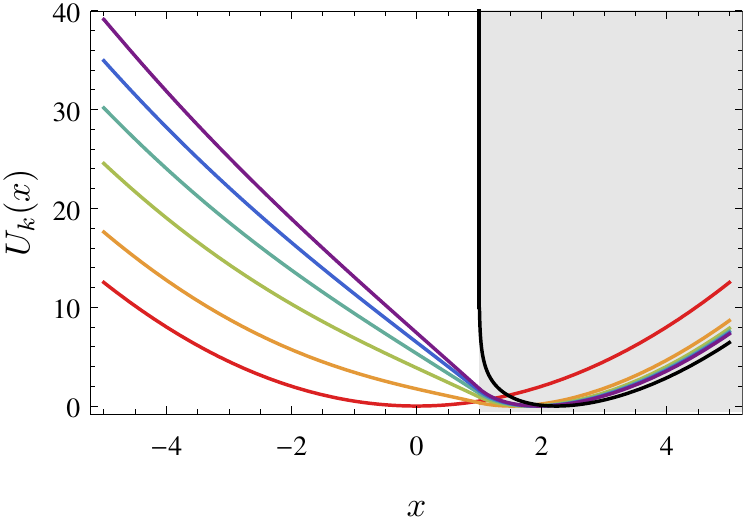}%
\hspace*{0.5in}%
\includegraphics[width=2.9in]{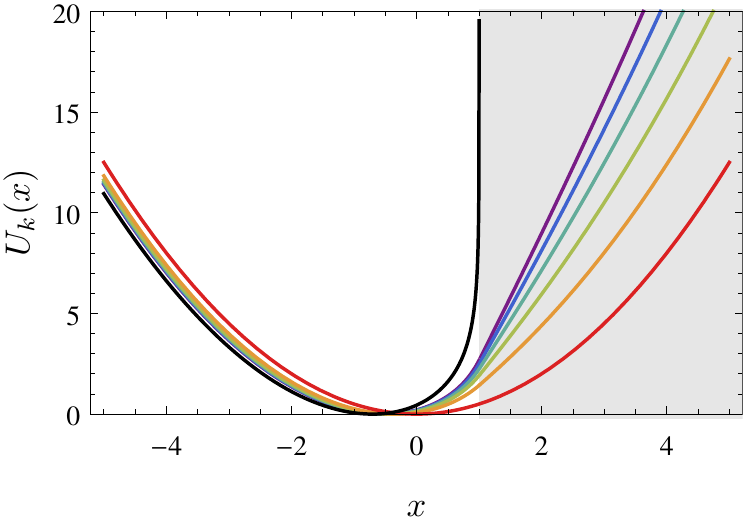}
\caption{(Color online) Effective potential $U_k(x)$ for $S=[1,\infty)$. Left: $k =0:2:10$ (from bottom to top colored curves) leading to more confinement in $S$. The black curve is the asymptotic effective potential $U_{\infty}(x)$ that confines the occupation of $X_t$ in $S$. Right: $k=0:-2:-10$ (from bottom to top colored curves) leading to less confinement. The black curve is the asymptotic effective potential $U_{-\infty}(x)$ preventing occupation in $S$. Parameters: $\alpha=1$, $D=2$.}
\label{fig:half:effective_potential}
\end{figure*}

Figure~\ref{fig:half:lambda} shows the results of the numerical calculation of $\lambda (k)$ and $I(r)$ from this matrix for $a=1$, which are overall qualitatively similar to those of Fig.~\ref{fig:rate} because of the restriction $r\in [0,1]$. In Fig.~\ref{fig:half:effective_potential} we show the effective potential $U_k(x)$ for positive and negative values of $k$ related to a confinement in the region $x>1$ and $x<1$, respectively. The shape of $U_k(x)$ is also qualitatively similar to the previous results obtained for $[0,1]$, except that it develops only one steep barrier instead of two. As before, the divergence of $U_k(x)$ near $x=1$ appearing in the limit $k\ra\pm\infty$ is logarithmic, since the wavefunction $\psi_k(x)$ of the quantum harmonic oscillator with a infinite wall at $x=1$, the equivalent quantum problem, vanishes at $x=1$ \cite{dean1966,consortini1976}. 

Another interesting feature to observe in Fig.~\ref{fig:half:effective_potential} is that the `natural' potential $U=U_0$ is not modified much by the conditioning on the side of occupation; that is to say, more occupation for $x>1$ (respectively, $x<1$) does not change the right (respectively, left) branch of $U$ significantly. This can be understood from the quantum perspective by noting that the introduction of a wall or well in the parabolic potential does not affect the tails of $\psi_k$ far away from this wall or well. The same phenomenon can also be explained using recent results \cite{chetrite2015} showing that the modified force $F_k$ minimizes a cost function involving a weighted integral of $(F-F_k)^2$ and the occupation $R_T=r$. As a result, the drift $F$ is modified only minimally whenever it contributes `naturally' to the occupation targeted. This cost, importantly, is a function of the drift and not the potential, so that large differences between $U_k$ and $U$, such as those seen in the right plot of Fig.~\ref{fig:asym:effective_potential}, do not necessarily translate into large differences between $F_k$ and $F$ and, therefore, large costs.

Considering $a=0$ instead of $a=1$ does not change these results much. The only difference is that the rate function shown in Fig.~\ref{fig:half:lambda} is symmetric about $r=0.5$, which leads to an effective potential $U_k(x)$ for $k<0$ that is the mirror image of $U_k(x)$ for $k>0$, that is, $U_k(x)=U_{-k}(-x)$. A logarithmic singularity near $x=0$ also appears for $a=0$ in the limits $k\ra -\infty$ and $k\ra\infty$, which restrict the occupation in the negative and positive regions, respectively. The positive case is interesting as it is related to the so-called arc sine law \cite{majumdar2005,rouault2002,kac1951} and leads to a generalization of the Brownian meander. Indeed, solving the Weber equation for $k\ra\infty$, which is equivalent to the quantum harmonic oscillator with a wall \cite{dean1966,consortini1976}, we find that the \emph{asymptotic Ornstein-Uhlenbeck meander} defined as the SDE (\ref{eq:ou1}) conditioned on staying positive \emph{at all times}, is a nonlinear diffusion with potential $U_{\text{OUm}}=U_\infty$ having the following tails:
\be
U_{\text{OUm}}(x)
\sim
\begin{cases}
-c \ln x & x\ra 0^+\\
\beta x^2/2 & x\ra\infty,
\end{cases}
\ee
where $c$ and $\beta$ are constants that can be determined numerically from $C(\lambda,k)$. The drift of this meander is thus given asymptotically by
\be
F_{\text{OUm}}(x)
\sim
\begin{cases}
c/x & x\ra 0^+\\
-\beta x & x\ra\infty.
\end{cases}
\ee
For pure Brownian motion ($\gamma=\alpha=0$), we find $\beta=0$, which is consistent with the exact drift of the Brownian meander; see Eq.~(21) of \cite{majumdar2015}.

\subsection{Point occupation}

\begin{figure*}[t]
\includegraphics[width=2.9in]{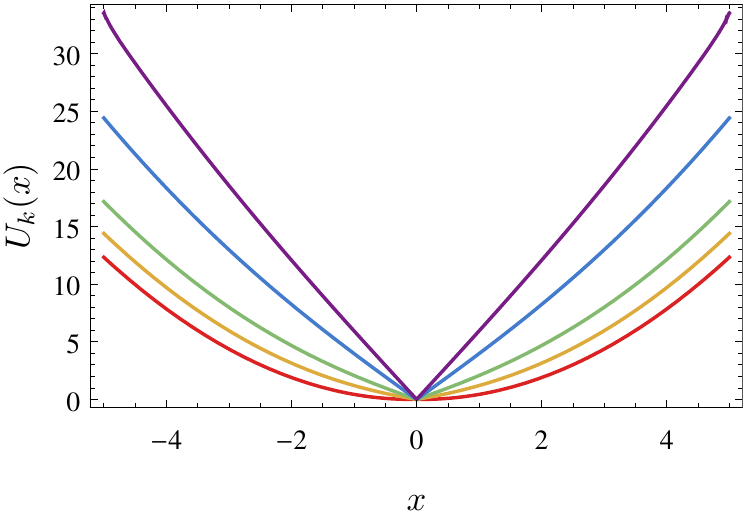}%
\hspace*{0.5in}%
\includegraphics[width=2.9in]{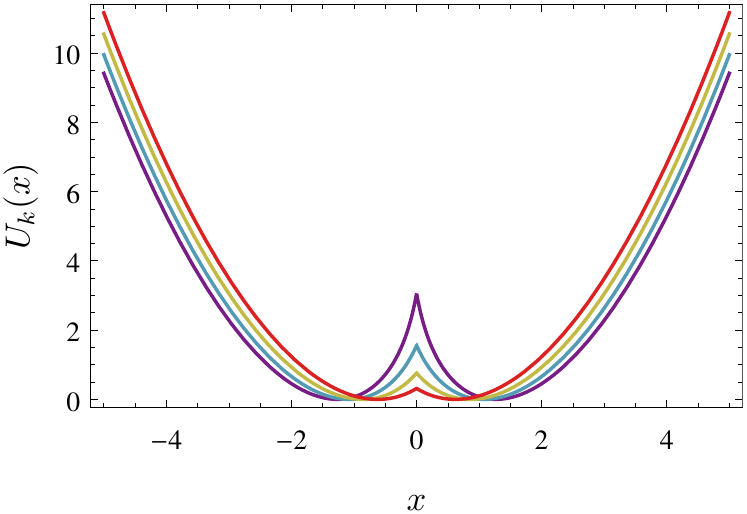}
\caption{(Color online) Effective potential $U_k(x)$ for the point occupation at $x=0$. Left: $k=0, 1.01, 2.02, 4.04$, and $6.06$ (from bottom to top curves) leading to more occupation at $x=0$. Right: $k=-1.01, -2.02, -4.04$, and $-10.1$ (from bottom to top curves at $x=0$) leading to less occupation at $x=0$. Parameters: $\alpha=1$, $D=2$.}
\label{fig:dirac:effective_potential}
\end{figure*}

The third and last application that we consider is the point occupation at $x=a$, obtained by replacing $\idf_S(x)$ by $\delta(x-a)$ in the definition of $R_T$ to obtain the local time at $a$. This case can also be considered as the limit $\epsilon\ra 0$ of $S=[a-\epsilon/2,a+\epsilon/2]$, with $\idf_S$ replaced by $\idf_S(x)/\epsilon$ and leads to the following Weber solution:
\be
\Psi (x) =  
\begin{cases} 
K_1 W( \nu ', -x) & x < a \\
K_2 W( \nu ', x) & x > a \\
\end{cases}
\ee
with continuity conditions
\be
\begin{aligned}
K_1 W(\nu ,-a) - K_2 W(\nu , a) &=0 \\
K_1 \partial_xW(\nu , -a) + K_2  \partial_xW(\nu ,a) & = k  \Psi (a).
\end{aligned} 
\label{eq:dirac:cont} 
\ee
In the particular case $a=0$, these conditions reduce to the following relation:
\be
 k = -\frac{\cot \p{\pi   \p{\frac {\nu } 2+\frac 1 4} } \Gamma  \p{\frac 1 4-\frac {\nu } 2}}{\sqrt {2}\, \Gamma  \p{ \frac 3 4-\frac {\nu } 2}},
\label{eq:dirac:zero}
\ee
which can be used as an implicit equation to find $\lambda(k)$ via the expression (\ref{eq:nu1}) of $\nu$.

The solution for $\psi_k$ that we find in this case is similar to the one found for the half line, except that the derivative of $\psi_k$ is now discontinuous because of the delta source at $x=a$, and jumps according to the second line in (\ref{eq:dirac:cont}). This introduces a kink at $x=a$ in the effective potential $U_k(x)$, illustrated in Fig.~\ref{fig:dirac:effective_potential} for $a=0$, which is reminiscent of the kink seen in the potential of Brownian motion with dry or solid friction \cite{gennes2005,touchette2010c,gnoli2013}. This makes sense intuitively: for the process $X_t$ to have a larger local time at $x=a$, it has to `stick' more onto that point, similarly to what is observed with solid friction. Conditioning on having a smaller local time at $x=a$ forces, on the other hand, $X_t$ to `avoid' that point as if there was a `negative' solid friction force (see left plot of Fig.~\ref{fig:dirac:effective_potential}). 

In the limit $k\ra\infty$, the potential $U_k$ becomes degenerate and concentrates the process on $x=a$, whereas for $k\ra-\infty$, it develops an infinite barrier at $x=a$ with two logarithmic branches that prevents occupation onto that point. The latter limit yields a $Q$-process version of the Ornstein-Uhlenbeck process conditioned on not reaching $x=a$, which also breaks ergodicity.

\section{Perturbation theory}

We complement the exact results of the previous section by developing a perturbation theory in the parameter $k$ for obtaining $\lambda(k)$, $I(r)$, and $U_k(x)$. In principle, this perturbation can be applied around any value $k$ for which the spectrum of $\cH_k$ or $\cL_k$ is known, even if $\cL_k$ is not symmetrizable \footnote{In this case, one has to use perturbation theory for non-self-adjoint linear operators.}. For simplicity we consider reversible processes with effective (self-adjoint) Hamiltonian $\cH_k$ and develop a perturbation in the form
\be\
\cH_{k+\Delta k} = \cH_k + \Delta k\, \idf _S.
\ee
A natural starting point is $k=0$, since $\cH_0=H$ is simply the Hamiltonian (obtained by symmetrization of $L$) of the quantum harmonic oscillator with shifted energy levels, so that $\lambda(0)=0$ and $\psi_0=e^{-U/2}=\sqrt{p^*}$.

The application of standard perturbation theory for self-adjoint operators with non-degenerate spectrum gives directly \cite{kato1995}:
\be
\partial _k \lambda _{n}(k) = \langle\Psi _{n}(k)|\idf_S |\Psi _{n}(k)\rangle
\label{eq:pert:evalue}
\ee
and
\be
\partial _k \Psi _{n}(k) = \sum _{m \ne  n} \frac{ \langle \Psi _{m}(k)|\idf_S |\Psi _{n}(k)\rangle  }{\lambda _{m}(k) - \lambda _{n}(k)} \Psi _{m}(k).\label{eq:pert:evector}
\ee
Here, we use the quantum bracket notation for the inner product, and now denote by $\lambda_n(k)$ and $\Psi_n(k)$ the $n$th eigenvalue of $\cH_k$ and its corresponding eigenfunction, respectively. The matrix elements
\be
N_{i,j}(k) =  \langle \Psi _{i}(k)|\idf_S |\Psi _{j}(k)\rangle
\ee
driving the `evolution' of $\lambda_n(k)$ and $\Psi_n(k)$ as a function of $k$ have a natural geometric interpretation: they represent an \emph{orthogonality defect} of the basis $\{\Psi_n(k)\}$ with respect to the modified inner product, 
\be
\langle  \Psi_m(k) |\idf_S | \Psi_n(k)  \rangle =\int_S \Psi_m^*(k,x) \Psi_n(k,x)dx,
\ee
which defines mathematically a semi-positive sesquilinear form. To complete these equations, we can calculate the evolution of the orthogonality defect matrix $N$ itself with the perturbation:
\newcommand{\Sym}[1]{\underset{#1}{\mathrm{Sym}}}   
\be 
\begin{aligned}
\partial _k N_{i,j}(k) &=  \langle \Psi _{i}(k)|\idf_S | \partial _k \Psi _{j}(k)\rangle  + \langle \partial _k \Psi _{i}(k)|\idf_S |  \Psi _{j}(k)\rangle \\
&=  \sum _{m \ne  i } \frac{ N_{i,m}(k) N_{m,j}(k) }{\lambda_i(k) - \lambda_m(k) } + \sum _{m \ne  j } \frac{ N_{j,m}(k) N_{m,i}(k) }{\lambda_j(k) - \lambda_m(k)},
\end{aligned}
\label{eq:N:ode}
\ee
which simplifies on the diagonal to
\be
\partial _k N_{i,i}(k) =  \sum _{m \ne  i}  \frac{2 |N_{i,m}(k)|^2} {\lambda_m(k) - \lambda_i(k) }.
\label{eq:ev:N:diag}
\ee
It can be checked that these differential equations for $N(k)$ admit the set of diagonal matrices as fixed points. Knowing that $k\ra-\infty$ is equivalent to total deconfinement in $S$, we find however that $N(-\infty)=0$ is the only stable fixed point reached in that limit. Similarly, since $k\ra\infty$ corresponds to total confinement in $S$, we must have $N(+\infty )= \id$.

Equations (\ref{eq:pert:evalue}), (\ref{eq:pert:evector}), and (\ref{eq:N:ode}) define a set of coupled nonlinear differential equations that can be used to find the SCGF $\lambda(k)$, which corresponds to the dominant eigenvalue $\lambda_0(k)$, and its associated eigenfunction $\Psi$, which corresponds to $\Psi_0(k)$. From the dominant eigenfunction, we then find the driven potential $U_k$ as in the previous section. Moreover, using the 0th component of (\ref{eq:pert:evalue}), we can obtain the rate function $I(r)$ by rewriting the parametric expression (\ref{eq:I:parametric}) as 
\be 
I\p{ N_{0,0} (k) } =  k N_{0,0} (k) - \lambda _0(k).
\label{eq:I:parametric:N}
\ee

\begin{figure*}[t] 
\includegraphics[width=2.9in]{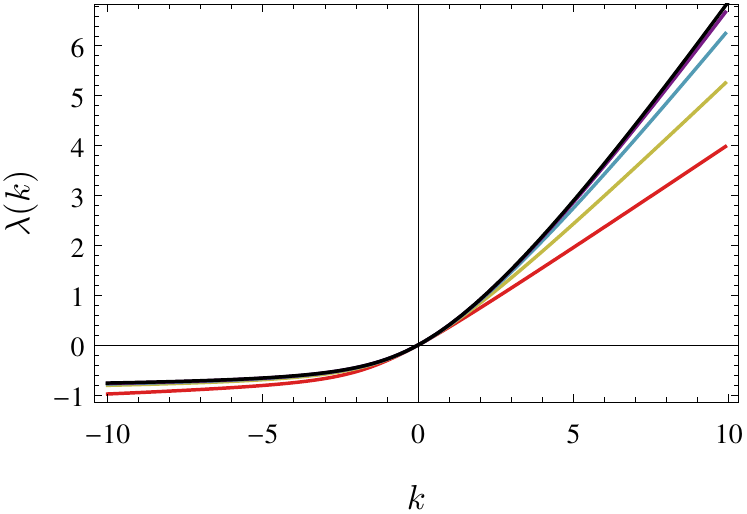}%
\hspace*{0.5in}%
\includegraphics[width=2.9in]{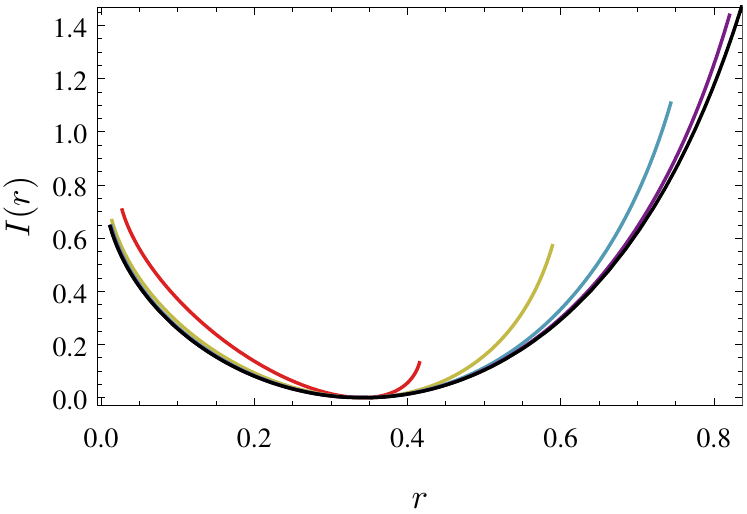} 
\caption{(Color online) Left: SCGF computed through perturbation for $S=[0,1]$ with $M=2,5,10,20$ modes (from bottom to top curves). The black curve shows  the exact $\lambda(k)$. Right: Corresponding rate function. The black curve shows the exact $I(r)$. Parameters: $\alpha=1$, $D=2$.}
\label{fig:rate:perturbative}
\end{figure*}

\begin{figure}[t]
\centering
\includegraphics[width=2.9in]{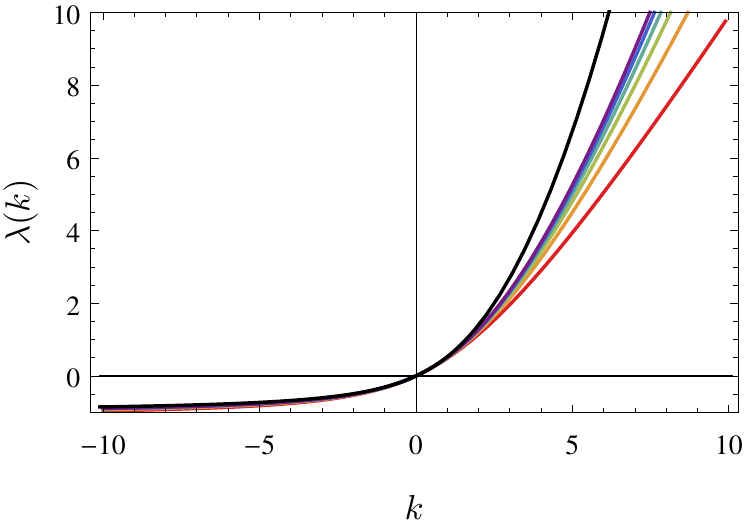}
\caption{(Color online) SCGF computed through perturbation for the point occupation at $x=0$. Number of modes used: $M=20:20:120$ (from bottom to top curves). The black line corresponds to the exact $\lambda(k)$. Parameters: $\alpha=1$, $D=2$.}
\label{fig:dirac:rate:perturbative}
\end{figure}

Figure~\ref{fig:rate:perturbative} shows the perturbation results for $\lambda(k)$ and $I(r)$ obtained by integrating Eqs.~(\ref{eq:pert:evalue}), (\ref{eq:pert:evector}), and (\ref{eq:N:ode}) starting from the known eigenvalues $\lambda_n(0)$ and eigenstates $\Psi_n(0)$ of the quantum harmonic oscillator. The results are for the unit interval occupation, $S=[0,1]$, and are also obtained by truncating $N(k)$ to a finite size $M$. As can be seen, the difference between the perturbative and exact results decreases by increasing $M$, as expected, and becomes negligible for $M=10$. The perturbation also converges quickly for $S=[1,\infty)$ (results not shown), but not for the point occupation case, as shown in Fig.~\ref{fig:dirac:rate:perturbative}. For the latter, the SCGF $\lambda (k)$ obtained by perturbation strongly differs from the exact result obtained from the methods of the previous section for $k$ beyond some positive value $k_c$, which is only slightly shifted by increasing $M$. This arises because the introduction of a Dirac \emph{well} in a potential (the quantum problem for $k>0$) strongly modifies the eigenfunctions and eigenvalues of that potential. By comparison, a Dirac \emph{wall} (the quantum problem for $k<0$) modifies these eigenfunctions essentially only in the way that they are joined at $x=a$, which leads to a small perturbation of the eigenvalues, including the dominant eigenvalue $\lambda(k)$ which converges to a constant as $k\ra-\infty$.

\begin{figure*}[t]
\includegraphics[width=\textwidth]{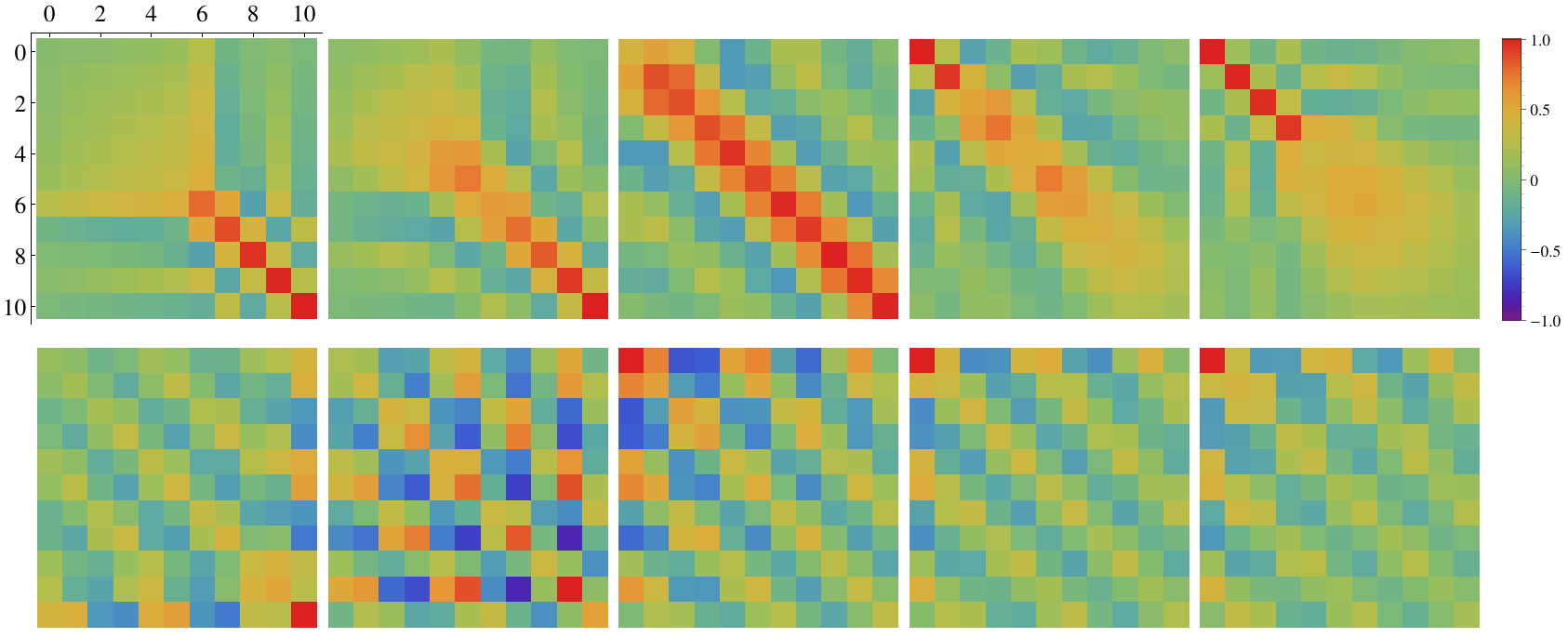}%
\caption{(Color online) Orthogonality matrix $N(k)$ for $k=-10,-5,0,5,10$ (from left to right). Top: $S=[1,\infty)$. Bottom: $S=[0,1]$. $M=10$ modes are used. Parameters: $\alpha=1$, $D=2$.}
\label{fig:Ns}   
\end{figure*}

To complete the perturbation analysis, we show in Fig.~\ref{fig:Ns} the evolution of the matrix $N(k)$ for increasing values of $k$, obtained by integrating the same differential equations truncated to order $M=10$. For $S=[1,\infty)$ (top panel), we see that this evolution essentially involves three phases: a first for $k<0$ in which $N$ has the approximate block form
\be
N\approx
\begin{pmatrix} 
0 & 0 \\
0 & \id
\end{pmatrix},
\label{eq:N:corners1}
\ee
a second around $k=0$ in which $N\approx\id$, and then a third phase obtained for $k>0$ in which 
\be
N\approx
\begin{pmatrix} 
\id & 0 \\
0 & 0
\end{pmatrix}.
\label{eq:N:corners2}
\ee
The first and last block phases are approximations of the extreme solutions $N(-\infty)=0$ and $N(\infty)=\id$, respectively, containing errors in the lower block coming from the truncation. In each case, the upper corner of $N$ follows the extreme solutions, which confirms that the largest eigenvalues~--~in particular, the dominant eigenvalue~--~are minimally affected by truncation. A similar evolution of $N(k)$ is observed for $S=[0,1]$ (lower panel), although convergence to $N(-\infty)=0$ and $N(\infty)=\id$ in this case is slower and involves more truncation errors outside the upper corner of $N$. These results are obtained with a direct truncation of $N(k)$ in the eigenbasis defined by $\Psi_n(k)$; a more efficient truncation leading to smaller errors could be constructed in principle by choosing a different function basis.

\section{Conclusion}

We have shown how a Markov process which is observed to spend a long time in some region of its state space can be represented by a modified Markov process, called the driven process, representing physically the dynamics of the original process restricted to that region. We have constructed this driven process for the Ornstein-Uhlenbeck process, and have shown how it can be used to obtain two important probabilistic constructions, namely, stochastic meanders which are confined in a certain region of space, and $Q$-processes which avoid a region of space.

The application of these results to higher-dimensional diffusions that are reversible should follow the example of the Ornstein-Uhlenbeck process. In this case, the driven process is obtained by solving a corresponding quantum ground state problem, as we have seen, which means that it can be solved using many powerful techniques of quantum mechanics (e.g., discretization, mesh or base function methods) \cite{thijssen2007}. For nonreversible diffusions, the problem is more complicated: there is no mapping to the quantum problem and the full spectral problem that must be solved involves, as mentioned, the tilted generator and its dual, with non-trivial boundary conditions imposed on the product of their respective eigenfunctions. An alternative method is to construct the driven process using optimal control representations detailed in~\cite{chetrite2015} or to discretize the underlying space to obtain a jump process which can then be studied using exact diagonalization or the density-matrix renormalization techniques developed in \cite{gorissen2009,hooyberghs2010,gorissen2011}.

For jump processes, the tilted generator becomes the tilted matrix
\be
\cW_k(x,y)= W(x,y)+k\idf_S(x)\delta_{x,y},
\ee
where $W(x,y)$ is the transition rate (probability per unit time) for the transition $x\ra y$, and $\delta_{x,y}$ is the Kronecker symbol. Moreover, the driven process is then the jump process with modified transition rates given by
\be
W_k(x,y)=r_k(x)^{-1} W(x,y) r_k(y),
\ee
where $r_k$ is the eigenvector associated with the dominant eigenvalue $\lambda(k)$ of $\cW_k$ \cite{chetrite2014}.

This result suggests many possible applications of the occupation conditioning problem beyond diffusions, including for example:
\begin{itemize}
\item Chemical reactions producing abnormally high or low concentrations of chemical species because of thermal noise. In this case, the state $X_t$ is the vector $(n^1_t,n^2_t,\ldots,n^m_t)$ of concentrations in time for $m$ chemical species so that $X_t\in\naturals_+^m$ or $X_t\in\reals_+^m$ \cite{kampen1992,gardiner1985}.

\item Queues in which the number $X_t\in\naturals_+$ of waiting `customers' goes beyond a certain threshold such as the queue capacity; see, e.g., \cite{iglehart1974,asmussen1982,shwartz1995}.

\item Random walks on regular or random graphs that visit `rare' or `metastable' nodes or graph components (e.g., nodes with low pagerank) \cite{montanari2002,kishore2012,kishore2013,bacco2015}. In this case, $X_t$ is simply the node visited at time $t$ while the state space is the set of nodes.

\item Interacting particle systems on lattice, such as the zero-range process, showing condensation transitions where a macroscopic number of particles get to occupy one lattice site \cite{grosskinsky2003,evans2005b,levine2005,grosskinsky2008}. The dynamics leading to this condensation and metastable phases related to it have been studied using occupation conditioning in \cite{grosskinsky2008,chleboun2010,chleboun2015}.

\item Other general Markov processes having metastable states; see, e.g., \cite{cassandro1984,beltran2010,bianchi2011} and references therein. The occupation set $S$ defining the conditioning can be chosen to include one or more metastable states or a set of states connecting stable and metastable states so as to study transition pathways, also called reactive paths. 
\end{itemize}

In all cases, the driven process provides a way to understand the dynamics of a stochastic process as it evolves in atypical states (concentrations, nodes, regions, etc.). This can take the form of a chemical reaction with modified rates, as already mentioned, or a queue with modified arrival and serving rates leading to a specific mean occupation. Similar interpretations apply to the other applications listed above, and should yield new insights in understanding in general how large fluctuations arise in time and how they can be simulated efficiently.

\appendix
\section{Large deviation principle for the occupation}
\label{appLDP1}

The contraction principle of large deviation theory states the following \cite{dembo1998,hollander2000,touchette2009}. Let $A_T$ be a random variable satisfying the LDP
\be
P(A_T=a)=e^{-TI_A(a)+o(T)}
\ee
with rate function $I_A(a)$ and let $B_T$ be another random variable such that $B_T=C(A_T)$. Then $B_T$ also satisfies an LDP,
\be
P(B_T=b)=e^{-T I_B(b)+o(T)},
\ee
with rate function
\be
I_B(b)=\min_{a:C(a)=b} I_A(a).
\label{eqcp1}
\ee
This is called the contraction principle because the function $C(a)$ can be many-to-one, in which case the fluctuations of $A_T$ are `contracted' to the fluctuations of $B_T$.

To use this result for the occupation $R_{T}$, as defined in (\ref{eqempmeas1}), we consider the empirical density
\be
\rho_T(x)=\frac{1}{T}\int_0^T \delta (X_t-x)\, dt
\ee
which represents the fraction of time (in the density sense) that $X_t=x$ over the time interval $[0,T]$. It is known from the Donsker-Varadhan theory (see \cite{dembo1998,hollander2000,touchette2009}) that the random function $\rho_T$ satisfies the LDP
\be
P(\rho_T=\rho) =e^{-TJ(\rho) +o(T)}
\ee
with the rate function given indirectly by the minimization shown in (\ref{eqdv1}). Given that
\be
R_{T}=\int_{\reals^m} \rho_T(x)\idf_S(x)\, dx =C(\rho_T),
\ee
we then obtain from the contraction principle (\ref{eqcp1}) the result shown in (\ref{eqcps1}).

\begin{acknowledgments}
We thank Satya Majumdar for useful comments. Financial support for this work was received from the National Institute for Theoretical Physics (Postdoctoral Fellowship), Stellenbosch University (Project Funding for New Appointee), and the National Research Foundation of South Africa (Grant nos 90322 and 96199).
\end{acknowledgments}

\bibliography{masterbib}

\end{document}